# An Agent-based Simulation of the Effectiveness of Creative Leadership

**Stefan Leijnen (stefan.leijnen@ubc.ca)**
Department of Psychology
University of British Columbia
Okanagan campus, 3333 University Way
Kelowna BC, V1V 1V7, Canada

**Liane Gabora (liane.gabora@ubc.ca)**
Department of Psychology
University of British Columbia
Okanagan campus, 3333 University Way
Kelowna BC, V1V 1V7, Canada

**Abstract**

This paper investigates the effectiveness of creative versus uncreative leadership using EVOC, an agent-based model of cultural evolution. Each iteration, each agent in the artificial society invents a new action, or imitates a neighbor's action. Only the *leader's* actions can be imitated by all other agents, referred to as *follower*s. Two measures of creativity were used: (1) *invention-to-imitation ratio*, $i_{Leader}$, which measures how often an agent invents, and (2) *rate of conceptual change,* $c_{Leader}$, which measures how creative an invention is. High $i_{Leader}$ increased mean fitness of ideas, but only when creativity of followers was low. High $i_{Leader}$ was associated with greater diversity of ideas in the early stage of idea generation only. High $c_{Leader}$ increased mean fitness of ideas in the early stage of idea generation; in the later stage it decreased idea fitness. Reasons for these findings and tentative implications for creative leadership in human society are discussed.

## Introduction

It is widely assumed that effective leaders are creative (Basadur, 2004; Bellows, 1959; Puccio, Murdock, & Mance, 2006; Simon, 1988; Sternberg, Kaufman & Pretz, 2003). Creativity, however, has drawbacks (Cropley, Cropley, Kaufman, & Runco, 2010). For example, a creative solution to one problem may generate other problems, and similarly, a creative solution to one element of a situation may have unexpected negative consequences with respect to other elements. Moreover, time spent creatively finding a solution for oneself is time not spent imitating and passing on solutions already found by others.

Previous investigations of the pros and cons of creativity using an agent-based simulation approach addressed the question: in an ideal society, what proportion of individuals should be 'creative types' (Leijnen & Gabora, 2009; Gabora, Leijnen & Ghyczy, in press)? The rationale was that in a group of interacting individuals only a fraction of them need be creative for the benefits of creativity to be felt throughout the group. The rest can reap the benefits of the creator's ideas by simply copying, using, or admiring them. After all, few of us know how to build a computer, or write a symphony or novel, but they are nonetheless ours to use and enjoy. Numerical simulations showed that if the proportion of creators is low, the mean fitness of ideas in the artificial society is highest when creators dedicate themselves fully to invention. However, as the proportion of creators increases, for optimal results, creators should spend more time imitating. Creative agents amounted to 'puncture points' in the fabric of society that interfered with the dissemination of proven effective ideas.

In the current investigation we focused exclusively on the extent to which creativity is desirable in a leader, where leadership is equated with having substantial influence over others. Previous results indicated that the presence of a leader accelerates convergence on optimal ideas, but does so at the cost of consistently reducing the diversity of ideas (Gabora, 2008b,c). In these previous simulations, the leader was no more nor less creative than the rest of the agents, referred to here as *followers*. The goal of the work reported here was to investigate how creative versus uncreative leadership affects the group as a whole.

## The Modeling Platform

Our investigation was carried out using an agent-based simulation referred to as 'EVOlution of Culture', abbreviated EVOC (Gabora, 2008b, 2008c). EVOC is an elaboration of Meme and Variations, or MAV (Gabora, 1994, 1995), the earliest computer program to model culture as an evolutionary process in its own right (as opposed to modeling the interplay of cultural and biological evolution). The approach was inspired by Holland's (1975) genetic algorithm, or GA. The GA is a search technique that finds solutions to complex problems by generating a 'population' of candidate solutions through processes akin to mutation and recombination, selecting the best, and repeating until a satisfactory solution is found. The goal here was to distil the underlying logic of not biological evolution but cultural evolution, i.e. the process by which ideas adapt and build on one another in the minds of interacting individuals. EVOC (as did MAV) uses neural network based agents that could (1) invent new ideas by modifying previously learned ones, (2) evaluate ideas, (3) implement ideas as actions, and (4) imitate ideas implemented by neighbors. Agents do not evolve in a biological sense—they neither die nor have offspring—but do in a cultural sense, by generating and sharing ideas for actions. EVOC (like MAV) successfully models how 'descent with modification' occurs in a cultural context. The approach can thus be contrasted with computer models of how individual learning affects biological evolution (Best, 1999, 2006; Higgs, 2000; Hinton & Nowlan, 1987; Hutchins & Hazelhurst, 1991).

EVOC consists of an artificial society of neural network based agents in a two-dimensional grid-cell world. It is written in Joone, an object oriented programming environment, using an open source neural network library written in Java. This section summarizes the key components of the agents and the world they inhabit.

## The Agent

Agents consist of (1) a neural network, which encodes ideas for actions and detects trends in what constitutes a fit action, and (2) a body, which implements actions.

**The Neural Network**. The core of an agent is an autoassociative neural network, as shown in Figure 1. It is composed of six input nodes that represent concepts of body parts (LEFT ARM, RIGHT ARM, LEFT LEG, RIGHT LEG, HEAD, and HIPS), six matching output nodes, and six hidden nodes that represent more abstract concepts (LEFT, RIGHT, FORELIMB, HINDLIMB, SYMMETRY and MOVEMENT). Input nodes and output nodes are connected to hidden nodes of which they are instances (e.g. RIGHT FORELIMB is connected to RIGHT.) Activation of any input node activates the MOVEMENT hidden node. Same-direction activation of symmetrical input nodes (e.g. positive activation—which represents upward motion—of both forelimbs) activates the SYMMETRY node.

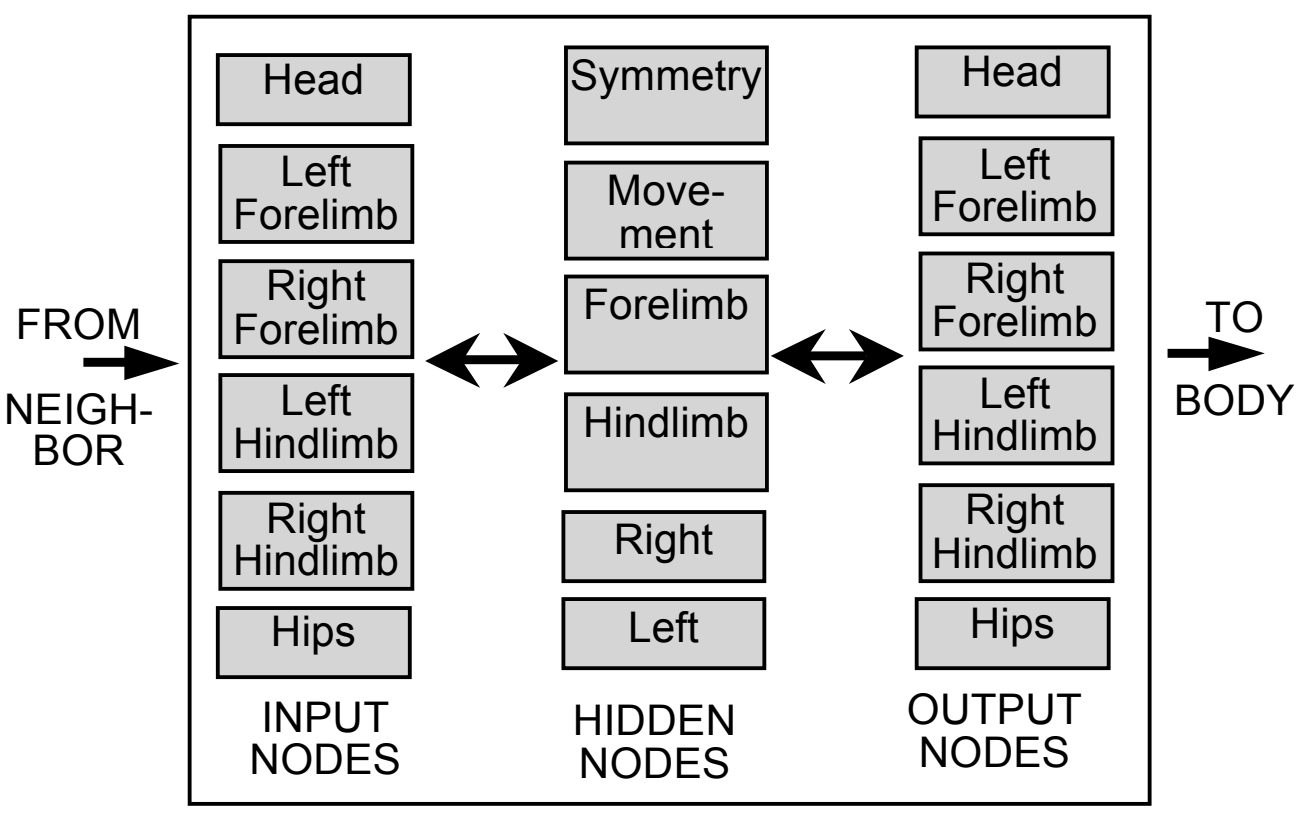

Figure 1. The neural network. See text for details.

The neural network learns ideas for actions. An idea is a pattern consisting of six elements that dictate the placement of the six body parts. Learning and training of the neural network is as per Gabora (1995). During imitation, the input is the action implemented by a neighbor. During invention, the pattern of activation on the output nodes is fed back to the input nodes, and change is biased according to the activations of the SYMMETRY and MOVEMENT nodes. In EVOC, the neural network can also be turned off to compare results with a data structure that cannot detect trends, and thus invents ideas merely at random.

**The Body**. If the fitness of an action is evaluated to be higher than that of any action learned thus far, it is copied from the input/output nodes of the neural network that represent *concepts of* body parts to a six digit array that contains representations of *actual* body parts, referred to as the *body*. Since it is useful to know how many agents are doing essentially the same thing, when node activations are translated into limb movement they are thresholded such that there are only three possibilities for each limb: stationary, up, or down. Six limbs with three possible positions each gives a total of 729 possible actions. Only the action that is currently implemented by an agent's body can be observed and imitated by other agents.

## The Fitness Function

Agents evaluate the effectiveness of their actions according to how well they satisfy needs using a pre-defined equation referred to as a *fitness function*. The fitness of an action with respect to the need to attract mates is calculated as in (Gabora, 1995). The fitness function rewards actions that make use of trends detected by the symmetry and movement hidden nodes and used by knowledge-based operators to bias the generation of new ideas. It generates actions that are relatively realistic mating displays, and exhibits a cultural analog of *epistasis*. In biological epistasis, the fitness conferred by the allele at one gene depends on which allele is present at another gene. In this cognitive context, epistasis is present when the fitness contributed by movement of one limb depends on what other limbs are doing.

## The World

MAV allowed only worlds that were square and toroidal, or 'wrap-around' (such that agents at the left border that attempt to move further left appear on the right border). Moreover, the world was always maximally densely populated, with one agent per cell. In EVOC the world can assume any shape, and be as sparsely or densely populated as required, with agents placed in any configuration. EVOC also allows for the creation of complete or semi-permeable permanent or eroding borders that decrease the probability of imitation along a frontier (although this was not used in the experiments reported here).

## Incorporation of Cultural Phenomena

Agents incorporate the following phenomena characteristic of cultural evolution as parameters that can be turned off or on (in some cases to varying degrees):

- **Imitation**. Ideas for how to perform actions spread when agents copy neighbors' actions. This enables them to share effective, or 'fit', actions.
- **Invention**. This code enables agents to generate new actions by modifying their initial action or a previously invented or imitated action. (See Gabora 1995 for further details.)

- **Knowledge-based Operators**. Since a new action (or, in invention, new idea for an action) is not learned unless it is fitter than the currently implemented action, new actions provide valuable information about what constitutes an effective idea. This information is used by knowledge-based operators to probabilistically bias invention such that new ideas are generated strategically as opposed to randomly. For example, if successful actions tend to be symmetrical (e.g. left arm moves to the right and right arm moves to the left), the probability increases that new actions are symmetrical. Also, if movement is generally beneficial, the probability increases that new actions involve movement of more body parts. (See Gabora 1995 for further details.)
- **Mental simulation**. Before committing to implementing an idea as an action, agents can use the fitness function to assess how fit the action would be if it *were* implemented.
- **Broadcasting**. Broadcasting allows the action of a *leader,* or broadcaster, to be visible to not just immediate neighbors, but all agents, thereby simulating the effects of media such as public performances, television, radio, or the internet, on patterns of cultural change. Each agent adds the leader as a possible source of actions it can imitate. The leader itself is thus the only agent that can only acquire actions from its immediate neighbors. The leader can be specified by the user or chosen at random. Broadcasting can be intermittent, or continued throughout the duration of a run. It can also be turned off altogether.

## A Typical Run

Each iteration, every agent has the opportunity to (1) acquire an idea for a new action, either by *imitation*, copying a neighbor, or by *invention*, creating one anew, (2) update the knowledge-based operators, and (3) implement a new action. To invent a new idea, for each node of the idea currently represented on the input/output layer of the neural network, the agent makes a probabilistic decision as to whether change will take place, and if it does, the direction of change is stochastically biased by the knowledge-based operators. If the new idea has a higher fitness than the currently implemented idea, the agent learns and implements the action specified by that idea. To acquire an idea through imitation, an agent randomly chooses one of its neighbors, and evaluates the fitness of the action the neighbor is implementing. If its own action is fitter than that of the neighbor, it chooses another neighbor, until it has either observed all of its immediate neighbors, or found one with a fitter action. If no fitter action is found, the agent does nothing. Otherwise, the neighbor's action is copied to the input layer, learned, and implemented.

Fitness of actions starts out low because initially all agents are immobile. Soon some agent invents an action that has a higher fitness than doing nothing, and this action gets imitated, so fitness increases. Fitness increases further as other ideas get invented, assessed, implemented as actions, and spread through imitation. The diversity of actions initially increases due to the proliferation of new ideas, and then decreases as agents hone in on the fittest actions.

## The Graphic User Interface

The graphic user interface (GUI) makes use of the open-source charting project, JFreeChart, enabling variables to be user defined at run time, and results to become visible as the computer program runs. The topmost output panel is shown in Figure 2.

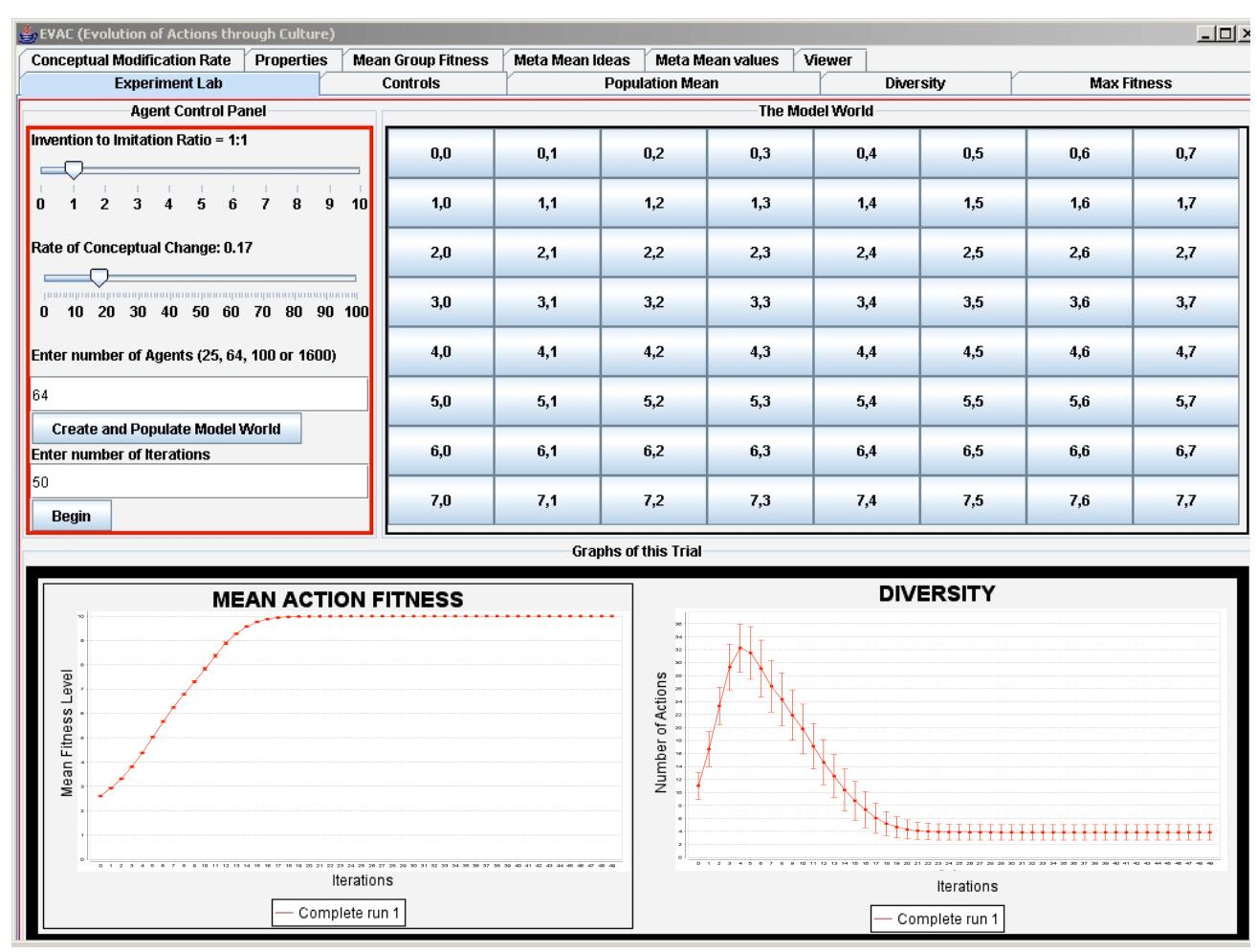

Figure 2. Output panel of GUI. See text for details.

At the upper left one specifies the *Invention to Imitation Ratio*. This is the probability that a given agent, on a given iteration, invents a new idea for an action, versus the probability that it imitates a neighbor's action. Below it is *Rate of Conceptual Change*, where one specifies the degree to which a newly invented idea differs from the one it was based on. Below that is *Number of Agents*, which allows the user to specify the size of the artificial society. Below that is where one specifies *Number of Iterations*, i.e. the duration of a run. Agents can be accessed individually by clicking the appropriate cell in the grid on the upper right. This enables one to see such details as the action currently implemented by that agent, or the fitness of that action. The graphs at the bottom plot the mean idea fitness and diversity of ideas. Tabs shown at the top give access to other output panels.

# Experiments

We now present the creative leadership experiments performed with EVOC. Unless stated otherwise, graphs plot the average of 100 runs, the world consists of 100 cells, one agent per cell, a 1:1 invention-to-imitation ratio, and a 1/6 probability of change to any body part during invention (The rationale behind this is that since, with six body parts, on average each newly invented action differs from the one it was based on with respect to one body part.)

The current experiments made use of EVOC's broadcasting function. As described above, broadcasting enables the action implemented by a leader to be visible throughout the artificial society. While experiments reported elsewhere investigated the impact of varying the number of leaders on the fitness and diversity of ideas (Gabora, 2008c), in the experiments reported here, simulated societies consist of one leader and ninety-nine followers. The leader is chosen randomly and broadcasts throughout the entire run.

## Experiment 1a: Effect of Varying Inventiveness (*i*) of Leaders and Followers on Fitness of Ideas

The first experiment investigated the effect of varying the ratio of iterations spent inventing versus imitating, or invention-to-imitation ratio, abbreviated *i*, of both the leader and the followers, on the fitness of ideas produced by the artificial society. The inventiveness of the leader, abbreviated $i_{Leader}$ was systematically varied from 0.0 to 1.0. When $i_{Leader}$ was 1.0, the leader invented a new action every iteration. When $i_{Leader}$ was 0.0, the leader never invented new actions; it only imitated its neighbors' actions. (It was still the leader because its actions were visible to, and could be imitated by, all other agents in the society, not just its immediate neighbors, as was the case for followers).

In the first set of runs, followers only imitated; they never invented, i.e., $i_{Followers}$ = 0.0. As shown in Figure 3, with uncreative followers, the degree of creativity of the leader matters a lot; the mean fitness of ideas in the artificial society is positively correlated with creativity of the leader.

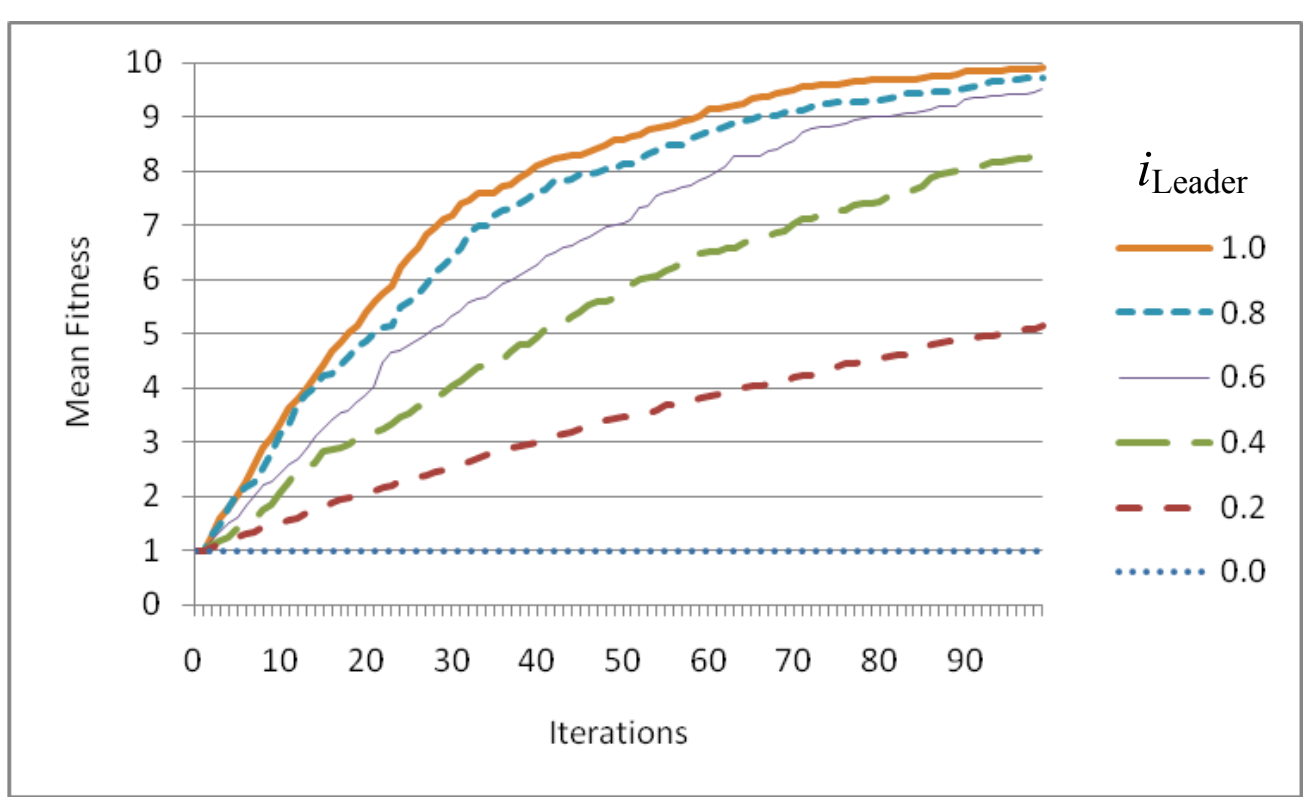

Figure 3. Mean fitness of actions with leaders of varying invention-to-imitation ratios, and followers that only imitate, i.e. that never invent (e.g. *i* = 0.0).

In the second set of runs, shown in Figure 4, followers were able to invent. More specifically, $i_{Followers}$ = 0.05; thus each iteration, each of the 99 followers had a 5% chance of inventing. Comparing figures 3 and 4 it is clear that while the degree of creativity of the leader had a large impact when followers are uncreative, it had almost no impact when followers were themselves creative. With creative followers, the mean fitness of ideas generated by the society increased over the duration of a run at more or less the same pace no matter how creative the leader was.

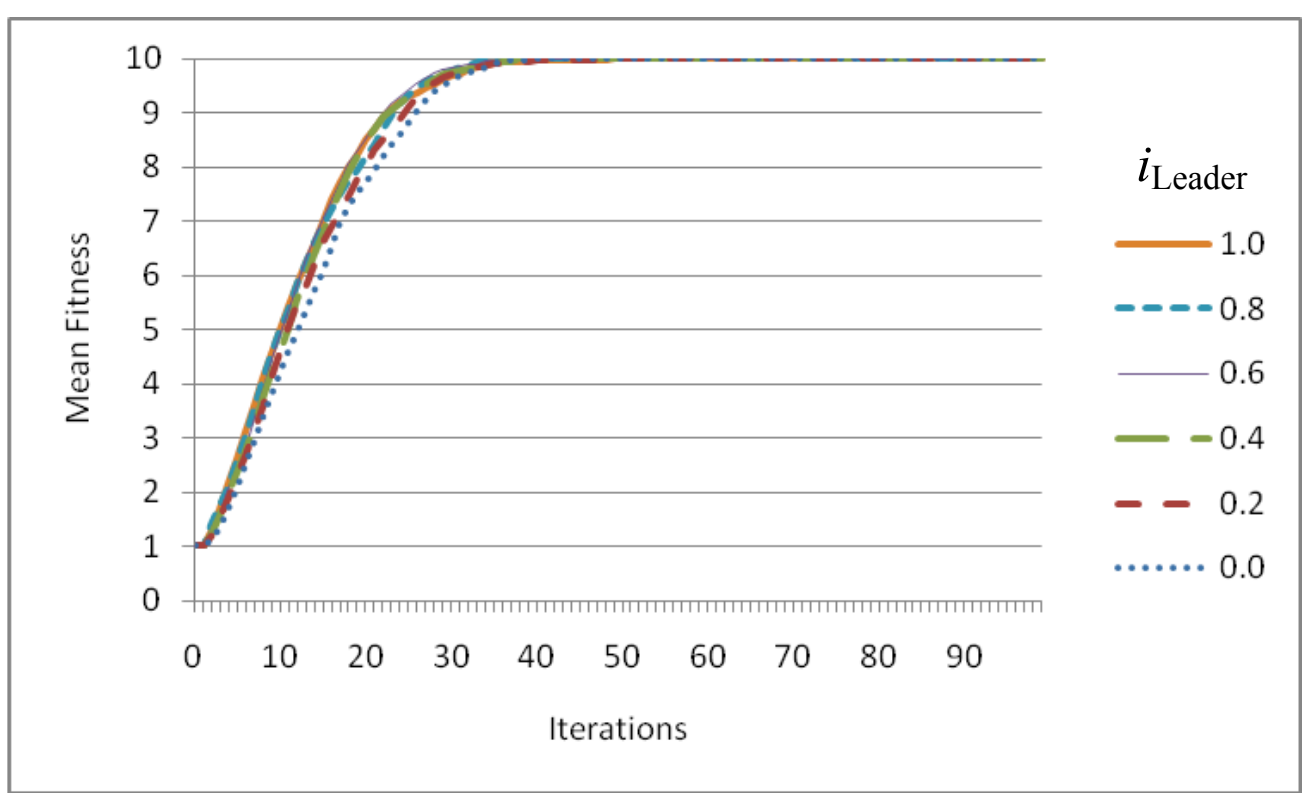

Figure 4. Mean fitness of actions with leaders of varying invention-to-imitation ratios, and followers that invent as well as imitate (*i* = 0.05).

## Experiment 1b: Effect of Varying Inventiveness (*I*) of Leaders and Followers on Diversity of Ideas

The second part of this experiment involved investigating the effect of varying the invention-to-imitation ratio, *i*, of both the leader and the followers on the *diversity* of ideas produced by the artificial society. As in experiment 1a, $i_{Leader}$ was systematically varied from 0.0 to 1.0. The result obtained with $i_{Followers}$ = 0.0 is shown in Figure 5.

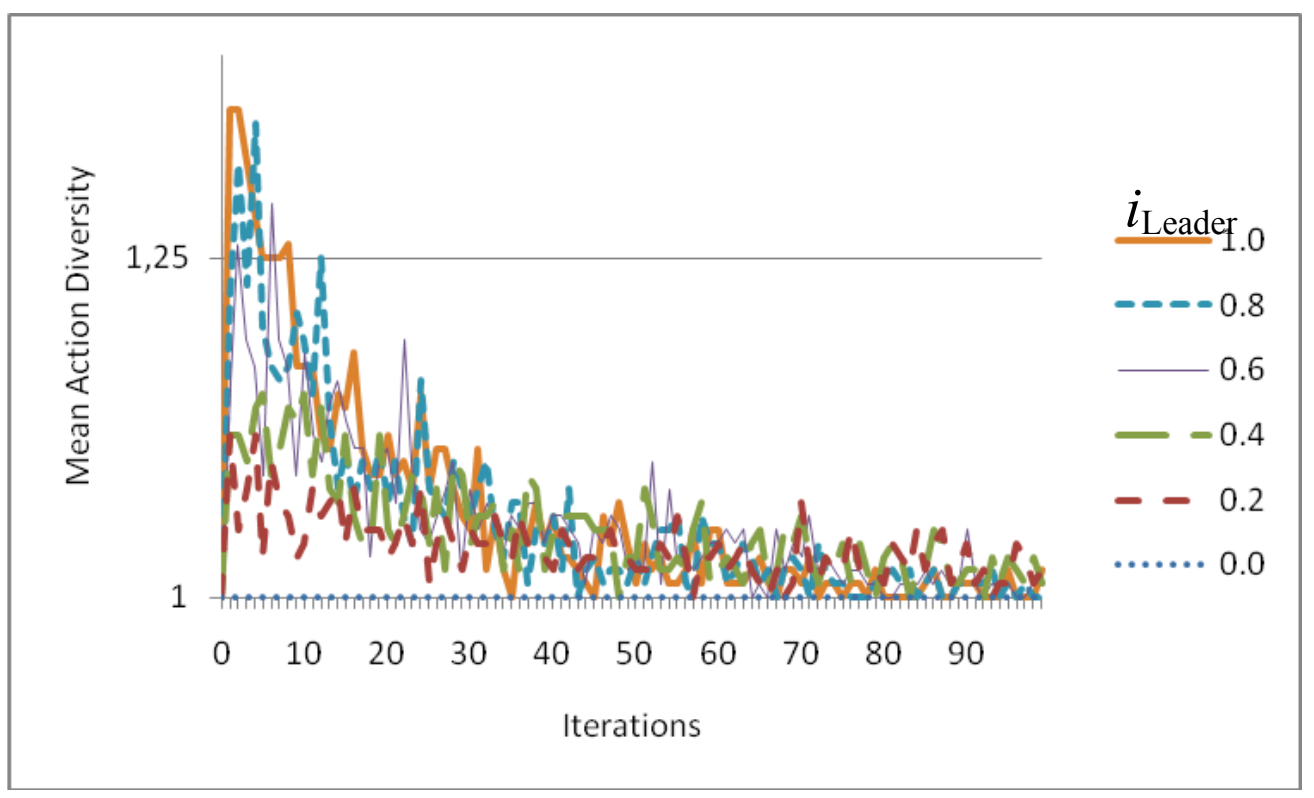

Figure 5. Diversity of actions in the artificial society with leaders of varying invention-to-imitation ratios, and followers that only imitate (*i* = 0.0).

In the short run, creative leadership was associated with increased diversity of actions. However in the long run, no matter how creative the leader, all agents converged on the same action, despite that there were seven other equally optimal actions they could have converged upon. Results with higher values of $i_{Followers}$ (not shown) were qualitatively similar. Action diversity was initially substantially higher, but it still always eventually converged to 1.

## Experiment 2: Effect of Varying Leaders' Rate of Conceptual Change (*c*)

There are two ways an agent's creativity can be manipulated in EVOC. The first way involves changing *i*, the invention-to-imitation ratio, as in the first set of experiments. It is possible to vary not just how frequently an agent invents, but how creative its newly invented ideas are. This second measure, referred to as the *rate of conceptual change,* abbreviated *c, is* implemented as follows. Invention occurs by taking the current action, and modifying it. When $c$ is low, the newly invented action varies little from the previous action upon which it was based. When $c$ is high, the newly invented idea varies dramatically from the previous idea upon which it was based.

As mentioned previously, the default value of $c$, the probability of change to any body part during invention, is 1/6 for any agent that invents, whether it is a leader or a follower. Previous experiments revealed this to be the rate that optimizes the rate of increase in mean fitness of actions (Gabora, 1995). Since ideas are ideas for actions, and since actions involve at most six body parts, on average, each newly invented action involves a change to the motion of one body part. Thus $c$ = 1/6 means that each body part changes what it is doing with a 1/6 probability, or 17% of the time. In this second set of experiments, shown in Figure 6, $c_{Leader}$ was systematically varied from 0% to 100%. Since the followers only imitated, $c_{Followers} = 0$. Because that means there are no new actions for the leader to imitate, $i_{\text{Leader}}$ was set to 1.0.

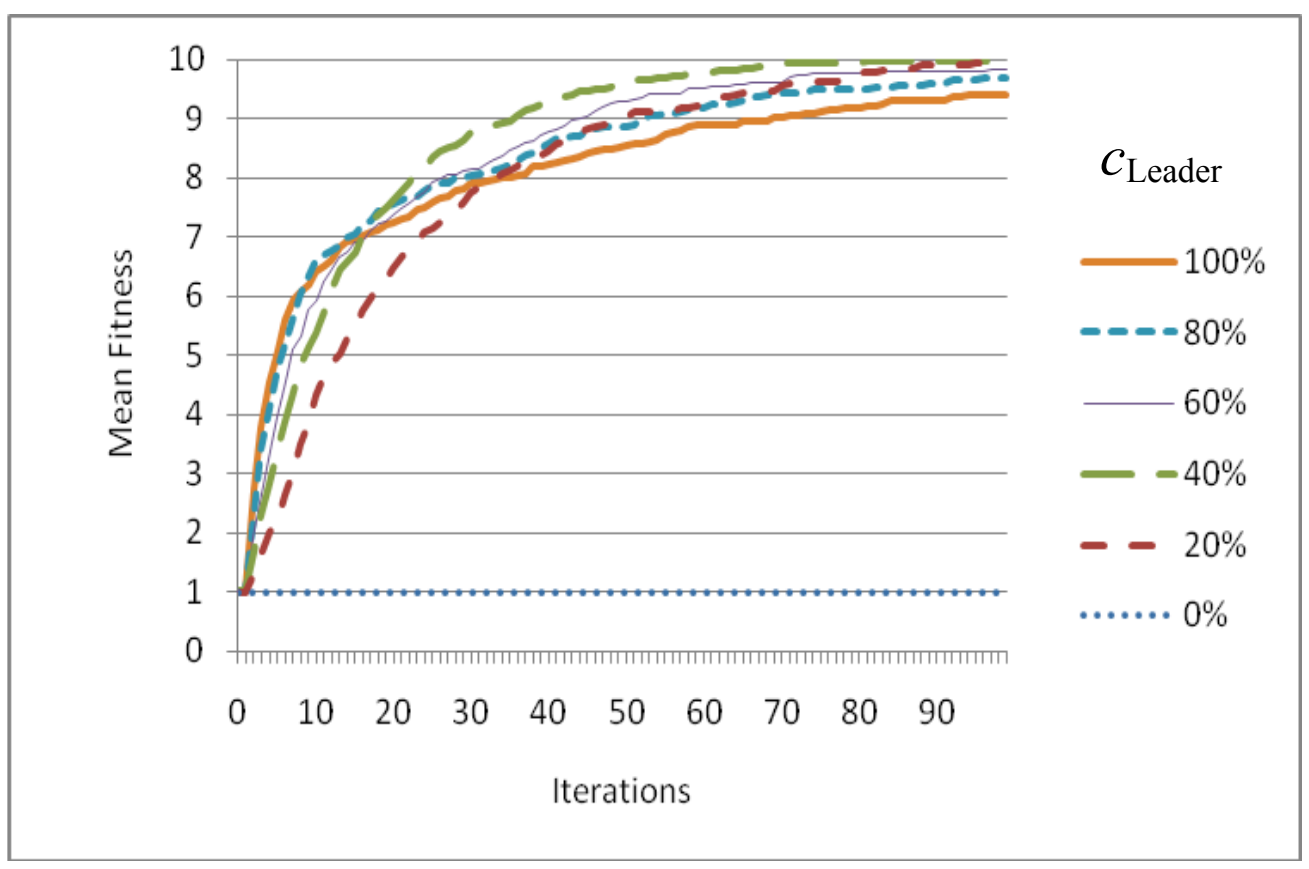

Figure 6. Mean fitness of actions in the artificial society with leaders of varying rates of conceptual change, and followers that only imitate.

Unlike in experiment one, the optimal degree of creative leadership with respect to this second measure of creativity depended on what phase of the creative process the society was at. Early on in a run, a form of leadership that entails the highest possible rate of conceptual change (100%) was most beneficial. However, as the run progressed a transition occurred, after which point a much lower rate of conceptual change (approximately 40%) was most beneficial.

## Discussion

The experiments reported here investigated the impact of creative versus uncreative leadership on the mean fitness and diversity of ideas for actions in an agent-based artificial society. The first experiment looked at the effect of varying the invention-to-imitation ratio of both leader and followers. The mean fitness of actions was positively correlated with the creativity of the leader, but only when the followers were uncreative. The more creative the followers, the greater the extent to which the beneficial effect of creative leadership was washed out. One must be cautious about extrapolating from a simple simulation such as this to the real world. For example, real-world creativity is correlated with emotional instability, affective disorders, and substance abuse (Andreason, 1987; Flaherty, 2005; Jamieson, 1993) which presumably would interfere with effective leadership, and which were not incorporated in these simulations. However, the result suggests that creativity may be a relatively unimportant quality for a manager of a creative team, but an important quality for a manager of an uncreative team.

The first experiment also investigated the effect of varying the invention-to-imitation ratio of both leader and followers on the diversity or number of different of actions implemented by agents. Previous results with EVOC had suggested that the beneficial effect of leadership on mean fitness of ideas is tempered by decreased diversity of ideas, and this echoed previous simulation findings that leadership can have adverse effects when agents can communicate (Gigliotta, Miglino, & Parisi, 2007). We wanted to know whether the decreased diversity associated with the presence of a leader was still observed when leaders are highly creative or highly uncreative compared to followers. We found that while in the early stages of a run, creative leadership (as well as the degree of creativity of followers) was associated with higher diversity, eventually all agents converged on what the leader was doing no matter how creative the leader (or how creative the followers). This suggests that in the long run leadership diminishes cultural diversity regardless of how creative the leader is. It is worth noting, however, that in this artificial world, unlike the real world, agents had only one task to accomplish. Further experiments will investigate whether these results hold true when the fitness function varies over time.

The second set of experiments investigated the effect of not how often the leader invents, but how creative any particular invention is, referred to as the rate of conceptual change. We found that early on in the creative process, when the fitness of the ideas that are getting generated was still relatively low, it was best if the leader was very creative (high rate of conceptual change). However, later in the creative process, once relatively fit ideas were being generated, a less creative leader was better (low rate of conceptual change). This result may reflect that the fitness function used here exhibits the cultural equivalent of the biological phenomenon of epistasis, wherein what is optimal for one element of an idea depends on what is going on with respect to another element. Initially, the higher the rate of conceptual change, the more quickly fitter actions are found. However, once relatively fit actions have been found, a high rate of conceptual change breaks up co-adapted epistatically linked elements and thus interferes with convergence toward optimal actions. In future experiments we will investigate whether these findings hold true when a different fitness function is used. However we believe that many real-world problem solving situations involve this kind of epistasis. Thus, although once again one must be cautious about extrapolating from the results of simple simulations such as this to the real world, our results suggest that a new startup company benefits most from highly creative leadership, while a more established company, or one that has stabilized on an established product line, benefits most from a more conservative form of leadership.

## Acknowledgments

This work is funded by grants to the second author from the Social Sciences and Humanities Research Council of Canada (SSHRC) and the GOA program of the Free University of Brussels.